%% file: la.tex
\documentclass{aa}

\usepackage{txfonts,graphicx,rotating,multirow}

\input makros
\begin{document}


\title{Condensation of MgS in outflows from carbon stars}


\author{Svitlana Zhukovska \and Hans-Peter Gail
}

\institute{Zentrum f\"ur Astronomie, Institut f\"ur Theoretische Astrophysik,
           Universit\"at Heidelberg, Albert-\"Uberle-Str. 2,
           D-69120 Heidelberg, Germany 
  }

\offprints{H.-P. Gail}

\date{Received date ; accepted date}

\abstract{}
{The basic mechanism responsible for the widespread condensation of MgS in the
outflows from carbon rich stars on the tip of the AGB is discussed with the aim of developing a condensation model that can be applied in model calculations of
dust formation in stellar winds.}
{The different possibilities how MgS may be formed in the chemical environment
of outflows from carbon stars are explored by some thermochemical calculations and by a detailed analysis of the growth kinetics of grains in stellar winds. The optical properties of core-mantle grains with a MgS mantle are calculated to demonstrate that such grains reproduce the structure of the observed 30 $\mu$m
feature. These considerations are complemented by model calculations of circumstellar dust shells around carbon stars.}
{It is argued that MgS is formed via precipitation on silicon carbide grains.
This formation mechanism explains some of the basic observed features of MgS
condensation in dust shells around carbon stars. A weak secondary peak at about 33 \dots\ 36~$\mu$m is shown to exist in certain cases if MgS forms a coating on
SiC. This new feature seems to have occasionally been observed.}
{}


\maketitle

\titlerunning{MgS grains from carbon stars}


\input sect1.lt

\input sect2.lt

\input sect3.lt

\input sect4.lt

\input sect5.lt

\input sect6.lt

\input lit.lt

\end{document}

%% file: makros.tex
\newcommand{\lapprox}{{\hskip 1pt}{\raise 1pt \hbox{$<$}}{\hskip-6.0pt}
{\lower 3pt \hbox{$\sim$}}{\hskip 2pt}}
\newcommand{\oder}[2]{{{\rm d}\,#1\over{\rm d}\,#2}}

%% file: sect1.lt
\section{Introduction}

Magnesium sulphide is a wide-spread dust component in dust forming stars with
carbon rich element mixture. Since Mg and S are abundant elements,
this dust species forms a major component in the dust mixture injected by 
dying AGB stars to the interstellar medium. Any attempt to model the
dust production by AGB stars and their contribution to the interstellar dust
population requires a knowledge of the production mechanisms at least of the
abundant dust species. Since for MgS a condensation model for this species is
presently lacking, we try to develop in this paper such a model. We will
explain, why we believe that MgS condenses in stellar outflows as a mantle on SiC
grains.

Magnesium sulphide was detected as dust component in circumstellar dust shells
enshrouding some carbon stars and in the environment of two planetary nebulae
by Forrest et al. (\cite{For81}) by observing a broad and prominent emission
band in the far-infrared spectrum which apparently is centred around 30
$\mu$m. At the instant of its detection the nature of the carrier of this
emission band was not immediately clear, except that the feature can only be a
solid state absorption band. Goebel \& Moseley (\cite{Goe85}) proposed MgS as
carrier of this feature. MgS has an infrared absorption feature just in the
wavelength region of the newly detected feature and chemical equilibrium
condensation  calculations of Lattimer et al. (\cite{Lat78}) for carbon rich
element mixtures, originally performed for identifying possible supernova
condensates, had shown that MgS should be an abundant solid phase at
temperatures of the order of 600 K or lower. A companion paper to the paper of
Goebel \& Moseley (\cite{Goe85}) by Nuth et al. (\cite{Nut85}) showed newly
determined optical properties of a number of solid phases, MgS, CaS, FeS,
SiS$_2$, FeS$_2$, and Fe$_3$C, that could be responsible for the 30 $\mu$m
band and which were compared to observations in the paper of Goebel \& 
Moseley. Though the match between the observed feature and the absorption band
of MgS was not perfect, the identification of MgS as carrier of the 30~$\mu$m
band has been widely accepted since that time, though other interpretations
are occasionally discussed (e.g. Papoular \cite{Pap00}; Grishko et al. 
\cite{Gri01}; Volk et al. \cite{Vol02}).

In order to allow to include a calculation of the MgS emission feature in
model calculations of radiative transfer in circumstellar dust shells, and
thus to enable a quantitative comparison between observations and theory,
Begemann et al. (\cite{Beg94}) determined optical constants of Mg$_x$Fe$_{1-x}$S
($0.9\ge x\ge 0$). They obtained good agreement between the observed emission
band profile and the calculated spectrum in the 30 $\mu$m region for a
radiative transfer model of the circumstellar dust shell of IRC+10216, which
left little doubt with respect to the interpretation of the 30$\mu$m emission
band as being due to MgS.

A number of observational and theoretical studies of the 30~$\mu$m emission
from dust enshrouded AGB and post-AGB stars, and from the environment of
planetary nebulae appeared since that time (e.g. Omont \cite{Omo93}; Omont et
al. \cite{Omo95}; Yamamura et al. \cite{Yam98}; Jiang et al. \cite{Jia99}; 
Szczerba et al. \cite{Szc99}; Hrivnak et al. \cite{Hri00}; Hony et al. 
\cite{Hon02}; Volk et al. \cite{Vol02}; Hony \& Bouwmann \cite{Hon04}; Lagadek
et al. \cite{Lag07}; Zijlstra et al. \cite{Zij06}; Leisenring et al. 
\cite{Lei08}). Good examples for the emission band can be found in Hony et al.
(\cite{Hon02}), Volk et al. (\cite{Vol02}) for galactic objects and Lagadek
et al. (\cite{Lag07}) and Zijlstra et al. (\cite{Zij06}) for the Magellanic clouds. They all show unequivocally that MgS formation is a common phenomenon
for carbon stars on the tip of the AGB. 

The formation mechanism of this MgS is not clear. It is already speculated in
the first papers by Nuth et al. (\cite{Nut85}) and Goebel et al. (\cite{Goe85})
that MgS forms via a surface reaction on carbon grains and this scenario was favoured over MgS condensation as a separate dust species. A labory investigation on MgS condensation was conducted by Kimura et al. (\cite{Kim05}), but this gives no direct insight in the formation process. No further discussion of this problem seems to have appeared in the astrophysical literature so far. In this paper we will show that MgS can only be formed in
the outflows of stars by precipitating on
pre-existing grains and that MgS in all likelyhood forms as mantle on 
the silicon carbide grains formed closer to the star at higher temperature.

The plan of the paper is as follows: First, we discuss some aspects of the
 chemistry associated with sulphur
bearing condensed phases in Sect. \ref{SectChem}. The kinetics of dust particle
growth in a stellar outflow is discussed in Sect. \ref{SectKinGr}. The results of some model calculations for MgS condensation in stellar winds are presented
in Sect. \ref{WindModels} and the optical properties of grains formed by MgS
grown on a silicon carbide or carbon core are briefly discussed in 
Sect.~\ref{SectCoreManle}. Our final conclusions are presented in Sect. \ref{SectConcl}.

%% file: sect2.lt
\begin{figure}[t]

\includegraphics[width=\hsize]{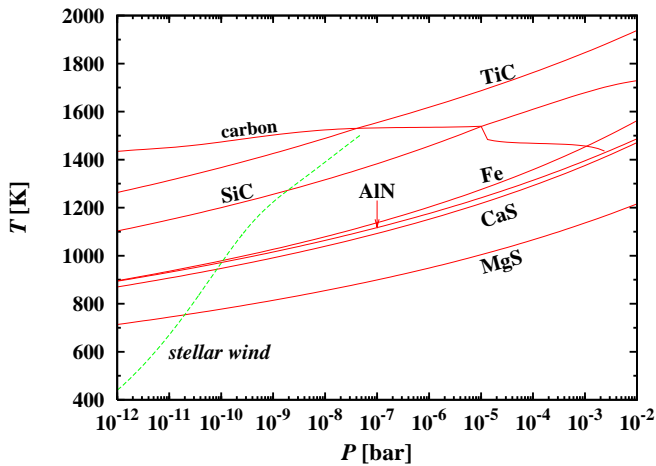}

\caption{Stability limits of solid phases formed from the most abundant elements
in chemical equilibrium for a carbon rich element mixture ($\epsilon_{\rm C}=
1.2~\epsilon_{\rm O}$). The dashed line shows the $P$-$T$-trajectory of a 
stationary dust driven stellar wind with a mass-loss rate of $\dot M=10^{-5}\,
\rm M_{\sun}\,yr^{-1}$}
\label{FigStabLim}
\end{figure}

\section{Stability of sulphur compounds}
\label{SectChem}

We start with a brief discussion of the equilibrium chemistry of sulphur
compounds in a carbon rich element mixture. Figure \ref{FigStabLim} shows the
results of a calculation of the chemical equilibrium composition of an element
mixture with cosmic element abundances (e.g. Grevesse \& Sauval\footnote{We
prefer these over the more recent data of Asplund et al. (\cite{Asp05}) since
the reduction of the C and O abundances in the new table are in clear
contradiction to the results of helioseismology (e.g. Chaplin et al. \cite{Cha07})} \cite{Gre98}) with carbon abundance set to 1.2 times the oxygen 
abundance. The lines in this $P$-$T$-diagram correspond to the upper stability 
limits of the indicated solid phases. For comparison we also show 
the wind trajectory of a stationary dust-driven wind model for the outflow from
a carbon star. In unconditional chemical equilibrium also some
oxygen bearing condensed phases would exist at temperatures below 800 K (cf.
Lattimer et al. \cite{Lat78}). The oxygen for their formation has to come from
a reduction of CO. It is unlikely that such reactions are kinetically possible
and we assumed here that the oxygen is completely blocked in CO which does not
react with other species in the system, i.e., we consider a conditional chemical
equilibrium where O is arrested in CO\footnote{Some condensation temperatures then are somewhat different from the results of more conventional condensation calculations}.

From the abundant elements with abundances relative to H exceeding $10^{-6}$, 
only CaS and MgS are formed as solid sulphur bearing phases in chemical
equilibrium. Troilite (FeS), that is abundant in oxygen rich environments, is
not formed in the carbon rich environment because MgS is more stable than FeS,
and in a carbon rich environment the Mg is not consumed by the even more
stable magnesium silicates as in the oxygen rich case. The first sulphur
bearing solid compound that appears in a cooling sequence at pressure
$10^{-10}$ bar is CaS at about 950 K, followed by MgS at about 750~K. The
temperatures where CaS and MgS become stable in a stellar outflow depend
somewhat on the mass-loss rate and are of the just mentioned order of
magnitude for a mass-loss rate of $\dot M=10^{-5}\,\rm M_{\sun}\,yr^{-1}$ or
somewhat less for smaller values of $\dot M$.

Calcium sulphide may be formed in a carbon rich environment prior to MgS
condensation, but because of the low calcium abundance (which is only 1/17
of the Mg abundance) it would be difficult to detect this dust component in
infrared spectra from circumstellar dust shells in the presence of MgS dust.
The formation of CaS cannot consume all the S since the sulphur abundance is
about eight times the Ca abundance, i.e., MgS is formed additionally to CaS at
lower temperatures. Since the Mg abundance is about twice that of the S 
abundance, at most half of the available Mg can be consumed by the formation of
MgS. The remaining fraction of Mg stays in the gas phase and may or may not
precipitate later on other condensed phases.

\begin{figure}[t]

\includegraphics[width=\hsize]{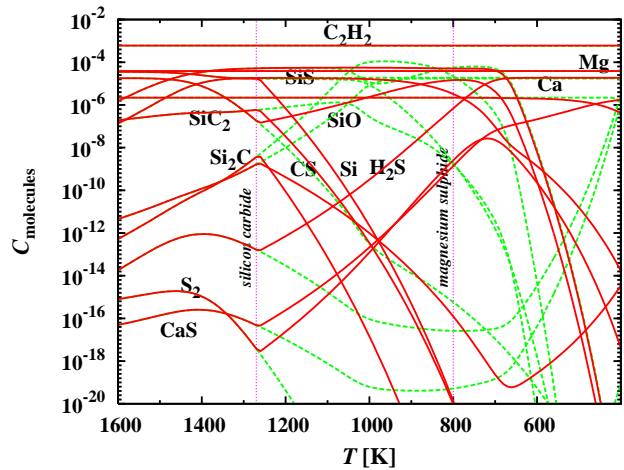}

\caption{Concentrations in the gas phase of S, Si, and Mg atoms and their
abundant molecular compounds with H, C, O, and S for a cooling sequence with
$P=10^ {-10}$ bar if formation of solid SiC in chemical equilibrium is allowed
for (full lines) and if the formation of SiC is suppressed (dashed lines)}
\label{FigChemSMol}
\end{figure}

The abundance of S bearing molecules in the gas phase is shown in
Fig.~\ref{FigChemSMol} for a cooling sequence with total pressure $10^{-10}$ bar
that is representative for the pressure in the dust formation zone of a stellar
wind (cf. Fig.~\ref{FigStabLim}). The figure shows the composition of the gas
phase in a restricted chemical equilibrium where carbon is not completely
condensed into solid carbon, a situation encountered in an outflow, where
carbon formation cannot keep pace with cooling and only part of the carbon is
converted from gaseous species into the condensed phase. Results for the
sulphur chemistry are shown for the two cases that (i) the Si condenses into
silicon carbide (SiC) at temperatures below about 1250 K and (ii) for the case
that formation of SiC is suppressed. A comparison of the results for both cases
shows:
\begin{itemize}
\item
If no SiC is formed, the sulphur is almost completely bound in SiS below 1500 K
and no other S bearing species are abundant in the gas phase. 
\item
Only if the Si is consumed by the formation of solid SiC, the sulphur is
liberated and forms increasing amounts of other S-bearing molecules -- in
particular H$_2$S -- in pace with conversion of Si from SiS into silicon
carbide.
\end{itemize}
This means that the sulphur chemistry of the gas phase and the formation of 
S-baring solid phases are strongly coupled to the process of silicon carbide
condensation.

As long as not much solid silicon carbide has formed the sulphur is blocked in
the rather stable SiS molecules and the formation of MgS from abundant gas phase
species requires the following reaction
\begin{equation}
\rm SiS\ +\ Mg\ {\buildrel k_1\over\longrightarrow}\ MgS(s)\ +\ Si\,.
\label{SiCfromSiS}
\end{equation}
The change of free enthalpy by this reaction is $\Delta G_1=-30\,\rm kJ/mol$ at 
800 K (thermochemical data are from Chase \cite{Chas98}). After the consumption
of SiS by silicon carbide formation and formation of abundant 
H$_2$S the following reaction becomes possible
\begin{equation}
\rm H_2S\ +\ Mg\ {\buildrel k_2\over\longrightarrow}\ MgS(s)\ +\ H_2\,.
\label{SiCfromH2S}
\end{equation}
The change of free enthalpy in this case is $\Delta G_2=-350\,\rm kJ/mol$. 
The corresponding laws of mass action for these two chemical reactions are
\begin{eqnarray}
a_{\rm MgS(s)}p_{\rm Si}&=&p_{\rm SiS}\,p_{\rm Mg}\,{\rm e}^{-\Delta G_1/RT}\\
a_{\rm MgS(s)}p_{\rm H_2}&=&p_{\rm H_2S}\,p_{\rm Mg}\,{\rm e}^{-\Delta G_2/RT}
\label{EqPH2S}
\end{eqnarray}
where $a_{\rm MgS(s)}$ is the activity of solid MgS. In chemical equilibrium
the activity of the solid compound equals unity. These equations determine the
equlibrium pressure in Sect.~\ref{SectDustGrowthEq}.

%% file: sect3.lt
\section{Formation of MgS in a stellar wind}
\label{SectKinGr}

In stellar outflows temperature and density variations occur on shorter
timescales than particle growth. In this case condensation proceeds far from
chemical equilibrium  and particle growth has to be treated by reaction
kinetics. We consider here  the conditions under which MgS condensation may
occur in a rapidly expanding stellar outflow.

\subsection{The wind model}

\subsubsection{Stationary wind model}

The formation of MgS in stellar outflows from AGB stars seems to be associated
with late phases of AGB evolution of low and intermediate mass stars, where the
mass-loss rate is very  high and the stars have lost already most of their
hydrogen rich envelope.

At the tip of the AGB, radiation pressure on dust becomes the dominating
driving source of the wind. The underlying stars seem all to be variables,
either Miras or LPVs. The structure of the outflow is complicated in this case,
since there are always several shocks running outwards, which are superposed
on the average outflow of stellar material. This is demonstrated by several
model calculations of dust forming stellar outflows of pulsating stars (e.g.
Bowen \cite{Bow88}; Fleischer et al.~\cite{Fle91,Fle92}; Feuchtinger et 
al.~\cite{Feu93}; H\"ofner \& Dorfi \cite{Hoe97}; H\"ofner et al.~\cite{Hoe98};
Winters et al.~\cite{Win97,Win00}; Jeong et al.~\cite{Jeo03}).

Calculating dust formation rates for multi-component dust mixtures of stars
with quite different elemental compositions and widely varying stellar
parameters presently seems not to be possible for reasons of computational time
requirements if the shock structure of the wind is to be taken into account.
For the purpose of calculating the composition of the dust mixture and the
amount of dust formed in the outflows of AGB-stars we use a more simple model,
which assumes a stationary outflow. If one compares the velocity and density
profiles of such stationary winds with published models of dust forming
pulsators (e.g. Winters et al.~\cite{Win00}; Jeong et al.~\cite{Jeo03}), one
observes in most cases a strong resemblance of the average velocity and density
profiles with that of the stationary models, except that in the pulsation
models outwards propagating shocks are superposed on the average flow
structure. So we can hope to obtain at least an estimate of the average
quantities of dust formed in the outflow if dust formation is calculated for
such an average outflow structure. One important consequence of this neglect of
the detailed structure of the flow is that we cannot determine
self-consistently the mass-loss rate $\dot M$. We have to treat this as a free
parameter, which has to be fixed in some other way.

\subsubsection{Stellar parameters}

The central star is assumed to be on the tip of the AGB. Its luminosity then is
typically $L_*=1.5\times10^4~\rm L_{\sun}$. 

For stars forming MgS in their outflow a rather high metallicity is required
since at low metallicities the abundances of Mg and S would be too low for this
to be possible. Only a small fraction of the carbon stars in the Small 
Magellanic Cloud, for instance, seem to form MgS dust (Sloan et al. 
\cite{Slo06}), while for carbon stars in the Large Magellanic Cloud the formation of MgS seems to be more common (Zijlstra et al. \cite{Zij06}), and in
the Milky Way it is quite common for highly evolved carbon stars (e.g. Hony
\cite{Hon02}). We note already at this place that according to observations MgS
formation seems to be associated with SiC formation: Stars showing the MgS
feature also show the SiC feature, but not all stars showing a SiC feature also
show the MgS feature in their spectra (e.g. Zijlstra \cite{Zij06}). In order
that low and intermediate mass stars of Pop I (or not much lower) metallicity
become carbon stars, their initial masses must be from the mass range between
about 1.5~M$_{\sun}$ and about  4~M$_{\sun}$.
 
After massive mass-loss on the AGB the stellar mass is likely to be already
substantially reduced and we assume in the model calculation a fixed stellar
mass of 1~M$_{\sun}$. The stellar radiation is approximated by a black body
spectrum. The effective temperature (in the inter-pulse phase) is determined
from the relation
\begin{equation}
\log T_{\rm eff}=.234\,\log M-.2\,\log L-.116\,\log{Z\over 0.02}+4.146
\end{equation}
given by Vassiliadis \& Wood (\cite{Vas93}) for stars on the AGB. The mass $M$
and luminosity $L$ are in solar units. With the assumed stellar parameters and
solar metallicity one finds $T_{\rm eff}=2\,250~\rm K$. 

The mass-loss rate is estimated from the relation
\begin{equation}
\dot M=2.1\times10^{-8}\ {L\over v_\infty}
\end{equation}
given by Vassiliadis \& Wood (\cite{Vas93}) for stars in the superwind phase.
The luminosity is in units of L$_{\sun}$, the outflow velocity $v_\infty$ in 
km\,s$^{-1}$, and the mass-loss rate in units of M$_{\sun}$\,yr$^{-1}$. With
the estimated stellar parameters and a typical outflow velocity in the
superwind phase of 10\,\dots\,15~km\,s$^{-1}$ one obtains a mass-loss rate of
about $\dot M=2\,\dots\,3\times10^{-5}\,\rm M_{\sun}\,yr^{-1}$ for stars at
the end of their AGB evolution. A value of $\dot M=3\times10^{-5}\,\rm
M_{\sun}\,yr^{-1}$ is used in the model calculations.

A typical run of density and temperature in a stationary stellar wind model
of a carbon star calculated for a dust driven stationary wind model (cf. Gail 
\& Sedlmayr \cite{Gai86,Gai87} for the details of how such models are
constructed) is shown in Fig.~\ref{FigStabLim}.

The C/O abundance ratio of the outflowing material in any case exceeds unity
since MgS is not formed in an oxygen rich environment.

\subsection{Dust particle growth}
\label{SectDustGrowthEq}

The model calculation of condensation in the outflow from a carbon star
considers the following dust components: Carbon dust, silicon carbide dust,
and iron dust. The formation of these dust species is calculated as outlined in
Ferrarotti \& Gail (\cite{Fer06}). Here we add the growth of magnesium sulphide
to our model. The dust formation is calculated by solving the equations for
dust growth (e.g. Gail \cite{Gai03})
\begin{equation}
\oder at=V_0\,\alpha\,n_{\rm gr}\,\sqrt{kT_{\rm g}\over2\pi m_{\rm gr}}\,\left\{
\Phi(w)-{p_{\rm eq} (T_{\rm d})\over n_{\rm gr}kT_{\rm d}}
\sqrt{T_{\rm d}\over T_{\rm g}}\ \ \right\}
\label{EqDuGrow}
\end{equation}
for all species of interest. Here $a$ is the radius of a dust grain, $V_0$ the
volume of one formula unit\footnote{
The formula unit corresponds to the chemical formula of the condensed phase, if
it is written with integer stoichiometric coefficients
}
of the chemical compound in the solid, $\alpha$ is
the growth coefficient, $n_{\rm gr}$ is the particle density of the growth
species in the gas phase, $m_{\rm gr}$ its mass, $p_{\rm eq}$ is the partial
pressure of the growth species in a state of chemical equilibrium between the
condensed phase and the gas phase, and $T_{\rm g}$ and $T_{\rm d}$ are the gas
temperature and the internal lattice temperature of the dust, respectively. The
quantity
\begin{equation}
\Phi=\sqrt{1+{\pi\mu m_{\rm H}\over8kT_{\rm g}}w^2}
\end{equation}
depends in principle on the drift velocity $w$ of the grains relative to the
gas. Here drift is neglected and we put $\Phi=1$ since for high mass-loss rates
drift velocities are small compared to the sonic velocity. The volume $V_0$ of
one chemical formula unit occupied in the solid is
\begin{equation}
V_0={A_{\rm d}m_{\rm H}\over\varrho_{\rm d}}\,,
\end{equation}
where $A_{\rm d}$ is the atomic weight of the chemical formula unit and 
$\varrho_{\rm d}$ the bulk density of the condensate.

\begin{table}

\caption{Basic data used for calculation of MgS grain growth}

\begin{center}
\begin{tabular}{l@{\hspace{1.cm}}rl}
\hline
\hline
\noalign{\smallskip}
Quantity & value & unit \\
\noalign{\smallskip}
\hline
\noalign{\smallskip}
$A$ & 56.37 & \\
$\varrho_{\rm d}$ & 2.68 & g\,cm$^{-3}$ \\
$V_0$ & $3.51\times10^{-23}$ & cm$^3$ \\
$\alpha$ & 0.2 & \\
$\epsilon_{\rm Mg}$ & $3.85\times10^{-5}$ & \\
$\epsilon_{\rm Si}$ & $3.58\times10^{-5}$ & \\
$\epsilon_{\rm S}$ & $1.85\times10^{-5}$ & \\
$\epsilon_{\rm d}$ & $10^{-13}$ & \\
$a_{\max}$ & 0.12 & $\mu$m \\
$v_{\rm th}$ & $1.52\times10^4$ & cm\,s$^{-1}$ \\
\noalign{\smallskip}
\hline
\end{tabular}
\end{center}

\label{TabDuGr}
\end{table}

The growth equation has to be complemented by an equation for the consumption
of the growth species from the gas phase. The details are described in the
paper by Ferrarotti \&  Gail (\cite{Fer06}) and are not repeated here.

As rate determining step for the formation of MgS via reaction 
(\ref{SiCfromH2S}) we assume the addition of a H$_2$S molecule from the gas
phase since this is the least abundant of the molecules involved in the
reaction. $p_{\rm eq}$ in Eq.~(\ref{EqDuGrow}) is calculated from from 
Eq.~(\ref{EqPH2S}) with activity $a_{\rm MgS}=1$. The thermodynamic data for
calculating $\Delta G$ are taken from JANAF-tables (Chase \cite{Chas98}).

The growth coefficient $\alpha$, the probability that a collision is followed
by a growth step, seems not to  be known. For the similar solid MgO with the
same structure and similar bonding properties Hashimoto (\cite{Has90}) measured
a growth coefficient of about 0.2 and we will use the same value for MgS.

\subsection{Conditions for grain growth}
\label{SectCondForGrowth}

Some insight into the origin of the difficulty to explain how MgS forms in the
outflow can be gained from considering a simplified growth model. We neglect 
in the growth equation (i) the vaporisation and (ii) the consumption of the
growth species by grain growth. Additionally we use in the expression for
the thermal velocity a representative value for the temperature in the
growth zone of the dust. The first neglect is generally not too bad for
dust growth problems since the vapour pressure $p_{\rm eq}$ usually rapidly
decreases with decreasing temperature below the stability limit of a solid.
The second neglect restricts the applicability of our considerations to the
early growth phase before the gas phase becomes exhausted of condensable
material. This initial growth phase is that which we now consider.

The particle density $n_{\rm gr}$ is the least abundant gas phase species
involved in the rate determining reaction for the growth. Its abundance is
usually some fraction $g$ of the least abundant element from which this gas
phase species is composed of. In our case, where reaction (\ref{SiCfromH2S}) is
assumed to be the key reaction for MgS growth, the least abundant growth 
species is H$_2$S and sulphur is the least abundant of the elements forming
this molecule. We write $n_{\rm gr}=g\epsilon_{\rm el}N_{\rm H}$, where 
$N_{\rm H}$ is the number density of H nuclei which can be calculated from the
mass density $\rho$ of the wind 
\begin{equation}
N_{\rm H}={\rho\over(1+4\epsilon_{\rm He})m_{\rm H}}\,.
\end{equation}
The mass density is given in a stationary outflow with mass-loss rate $\dot M$ and
outflow velocity $v$ by
\begin{equation}
\rho={\dot M\over 4\pi r^2 v}\,.
\end{equation}

Since condensation of MgS becomes possible only at a temperature much lower
than the condensation temperature of carbon dust, the radiation pressure on
carbon dust has driven the outflow velocity already nearly to the terminal
outflow velocity of the stellar wind if MgS condensation commences. For the
problem of MgS condensation we can consider the wind velocity $v$ therefore to
be constant. By changing the independent variable $t$ to $r$ by means of $vdt=
dr$ we can write the equation for the initial growth of dust as
\begin{equation}
\oder ar=V_0\,\alpha\,v_{\rm gr}\,{\dot M\over1.4m_{\rm H}\,4\pi r^2 v^2}\,.
\end{equation}
Here $v_{\rm gr}$ is the thermal velocity of the growth species (assumed to
be constant). Integrating from the inner radius $R_{\rm d}$ of the onset of
growth of the considered dust species to infinity we obtain
\begin{equation}
a_\infty=a_0+V_0\,\alpha\,v_{\rm gr}\,g\epsilon_{\rm el}\,
{\dot M\over1.4m_{\rm H}\,4\pi R_{\rm d}\,v^2}\,,
\end{equation}
where $a_0$ is the radius of the seed nuclei for dust growth and $a_\infty$ the
radius at infinity. This equation requires that the grain radius $a_\infty$
remains smaller than the maximum radius $a_{\max}$ attained if all condensable
material is condensed. This maximum possible radius is given by
\begin{equation}
{4\pi\over3}a_{\max}^3\epsilon_{\rm d}={V_0\epsilon_{\rm key}\over\nu_{\rm key}}\,.
\label{DefMaxVolCond}
\end{equation}
Here $\epsilon_{\rm d}$ is the number of dust grains per hydrogen nucleus,
$\epsilon_{\rm key}$ the key element for the formation of the dust species
considered, and $\nu_{\rm key}$ the number of atoms of the key element in
the chemical formula of the solid.

We can now write
\begin{equation}
a_\infty=a_0+a_{\max}\,{\dot M\over\dot M_{\rm cr}}
\end{equation}
with
\begin{equation}
\dot M_{\rm cr}={a_{\max}\,1.4m_{\rm H}\,4\pi R_{\rm d}\,v^2\over
V_0\,\alpha\,v_{\rm gr}\,g\epsilon_{\rm el}}\,.
\end{equation}
The quantity $\dot M_{\rm cr}$ defines a critical mass-loss rate for particle
growth. If the mass-loss rate $\dot M$ of the stellar outflow is lower than
$\dot M_{\rm cr}$, the final particle radius of dust grains of the considered 
kind in the outflow is \emph{smaller than the maximum radius} attained in case
of complete condensation. Condensation by growth starting with some kind of
small seed particles is always incomplete in this case and most of the 
condensable material remains in the gas phase. If, however, the mass-loss rate
$\dot M$ of the stellar outflow is of the order of $\dot M_{\rm cr}$ or exceeds
$\dot M_{\rm cr}$, the final particle radius in the outflow approaches 
$a_{\max}$; complete condensation is possible in this  case by growth starting
with some kind of small seed particles (at least this is not kinetically
forbidden).

Applying this to MgS condensation we obtain
\begin{equation}
\dot M_{\rm cr, MgS}=5.4\times10^{-4}~\rm M_{\sun}\,yr^{-1}\,.
\end{equation}
Here we assumed a typical outflow velocity of $v=10~\rm km\,s^{-1}$ and a radius
of $R_{\rm d}=2\times10^{14}~\rm cm$ corresponding to about 5 stellar radii.
As key element for MgS condensation we assume sulphur since it has lower
abundance than Mg. The growth species is H$_2$S. As typical temperature for
calculating $v_{\rm gr}$ we choose 600~K. All other quantities are given in
Table~\ref{TabDuGr}. The critical mass-loss rate is about a factor of ten
higher than the highest mass-loss rates observed for highly evolved AGB stars
(e.g. van Loon et al. \cite{vLo99}). As a result the radius can approach at
most 1/10 of that for complete condensation and only about 1/1000 of the
condensable material can be condensed into MgS grains if the dust grains start
their growth for instance from seed particles with sizes of the order of 1~nm
that are typical sizes for seed particles resulting from nucleation from the gas
phase. 

For such a low fraction of condensation of S and Mg into MgS this dust
component would not be detectable in the infrared emission from circumstellar
dust shells of AGB stars. Since MgS is frequently observed in highly evolved
AGB stars, it cannot be formed by condensation from the gas phase via nucleation
and subsequent growth. We need therefore to look for some other kind of
formation mechanism. 

\subsection{Heterogeneous growth}

The most promising alternative is heterogeneous growth on some other
kind of dust particles that have already not-too-small radii at the onset of
MgS growth in order that the frequency of collisions of such a particle with
growth species from the gas phase is high already from the beginning on. We
derive the condition for this type of growth to occur by considering the initial
growth phase, where the growth of a coating of MgS on some type of carrier grain
has not yet significantly increased the total grain radius. 

We use essentially the same type of simplifications as before and consider the
following equation for the increase of the volume $V$ of the coating on the
surface of carrier grains of radius~$a_{\rm c}$
\begin{equation}
\oder Vt=V_0\alpha v_{\rm gr}n_{\rm gr}\,4\pi a_{\rm c}^2\,.
\end{equation}
Introducing again $r$ as independent variable we obtain upon integrating this 
equation from the radius $R_{\rm d}$ of the onset of formation of the coating
to infinity
\begin{equation}
V_\infty=V_0\alpha v_{\rm gr}\,4\pi a_{\rm c}^2\,g\epsilon_{\rm el}\,
{\dot M\over1.4m_{\rm H}\,4\pi R_{\rm d}\,v^2}\,.
\label{VolCoat}
\end{equation}
This is the final volume of the coating at large distances. The maximum
possible volume $V_{\max}$ of the coating on a carrier dust grain is, again,
given by Eq.~(\ref{DefMaxVolCond}). We can now write
\begin{equation}
V_\infty=V_{\max}\,{\dot M\over\dot M_{\rm het}}\,,
\end{equation}
where
\begin{equation}
\dot M_{\rm het}={\epsilon_{\rm key}\over\nu_{\rm key}\epsilon_{\rm d}}\
{1.4\,m_{\rm H}R_{\rm d}v^2\over\alpha v_{\rm gr}\,a_{\rm c}^2\,g
\epsilon_{\rm el}}
\end{equation}
is a characteristic mass-loss rate of the stellar outflow for formation of
coatings on the surface of already existing grains. If the mass-loss rate
$\dot M$ of the stellar outflow is less than the characteristic mass-loss rate
$\dot M_{\rm het}$ only part of the condensable material will condense in form
of a coating on the surface on some carrier grains.  If the mass-loss rate
$\dot M$ of the stellar outflow exceeds the characteristic mass-loss rate
$\dot M_{\rm het}$ the volume of the coating approaches the maximum possible
volume and complete condensation is possible, at least in principle. 

\begin{figure}[t]

\includegraphics[width=\hsize]{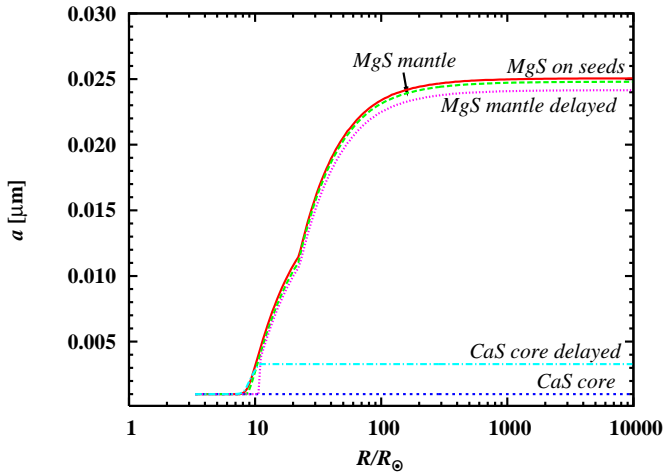}

\caption{Evolution of MgS and CaS grain radius as function of distance from the
centre of the star, for MgS formation (i) on seed nuclei of 1~nm size, and (ii)
as mantle on a CaS core. Also shown are the results of a model with delayed 
onset of MgS condensation on CaS core; details are given in the text. }
\label{FigMgS_CaS}
\end{figure}

Inserting numerical values for MgS condensation and assuming that the carrier
grains for the formation of the coating have a radius of $a_{\rm c}=0.1~\mu$m
one obtains
\begin{equation}
\dot M_{\rm het}=2.8\times10^{-4}~\left(0.1\,\mu{\rm m}\over a_{\rm c}\right)^2
~\rm M_{\sun}\,yr^{-1}\,.
\end{equation}
This characteristic mass-loss rate, again, is about a factor of ten higher than
the highest mass-loss rates observed for AGB stars, but since it now refers
to the volume we have the result, that at the end of the AGB evolution up to
about 10\% of the condensable material for MgS formation can really be
condensed  as coating on suited carrier grains, if such exist. If the core
grains have a bigger radius than the assumed 0.1~$\mu$m, or if the mass-loss
rate should be exceptionally high the fraction of condensed material may be
even bigger. Such amounts of condensed MgS should be clearly visible in the
infrared spectrum.

Hence, we come to the conclusion that the observed MgS in the outflows from
carbon stars on the AGB does not form as a separate dust species but condenses
as coating on the surface on some different kind of carrier grains. The
formation of significant amounts of MgS dust requires mass-loss rates of at
least a few times $10^{-5}~\rm M_{\sun}\,yr^{-1}$, i.e., this is possible only
during the superwind phase short before termination of AGB evolution if the
mass-loss rates have grown to very high values. This is just what one seems to
observe.

As a side-remark we note the following: Since the available amount of S does
not suffice for the formation of thick coatings of MgS such that the final
volume is much bigger than the volume of the carrier grains, the volume of
condensed material on the surface of the carrier grains increases quadratically
with the radius of the carrier grains (cf. Eq.~\ref{VolCoat}). Therefore it may
well be that most of the MgS is concentrated on a sub-population of 
lager-than-the-average grains, if one has a broad distribution of grain sizes
for the possible carrier grains. 

\begin{figure}[t]

\includegraphics[width=0.9\hsize]{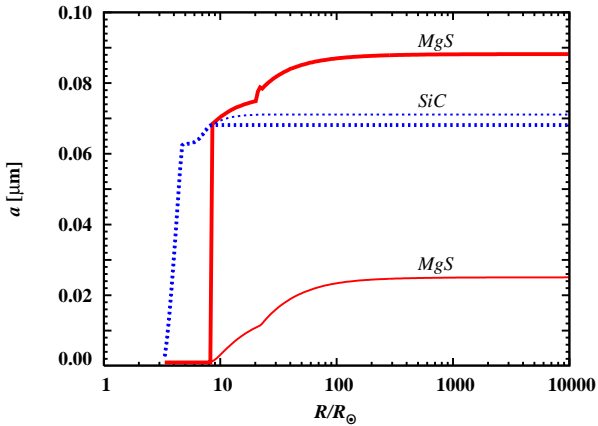}

\caption{Evolution of radius of SiC and MgS grains as predicted (i) by the model
for MgS condensation on seed nuclei, and (ii) as mantle on a SiC core (thin and
thick lines, respectively).}
\label{FigMgS_SiC}
\end{figure}

\subsection{Possible carrier grains}

The dust species that may serve as carrier grains for MgS formation need to
form dust grains with a size not much smaller than about $a_{\rm c}=0.1~\mu$m
because otherwise the condensed fraction of MgS becomes too small if the radius
of the cores is smaller than this. This restricts the possibilities, which dust
species may serve as substrates of MgS growth, to the most abundant dust species.

The outflowing gas enters the temperature regime below about 650 K favourable
for MgS condensation with two major dust species that have grown farther inside
in the shell to the required size: SiC grains and carbon grains. From 
investigations of presolar dust grains it is 
known that from the many poly-types of SiC only two are formed in circumstellar
environments (Daulton et al. \cite{Dau03}): The cubic 3C poly-type (usually
denoted as $\beta$-SiC) is found in 79.4\,\% of all observed grains, and the
hexagonal 2H poly-type (usually denoted as $\alpha$-SiC) is found in 2.7\,\% 
of all observed grains. In 17.1\,\% of all grains the material is an
intergrowth of these two poly-types.

The lattice structure of the SiC grains in stellar outflows, thus, in most cases
is cubic. Magnesium sulphide also has cubic lattice structure. The
electronegativity difference (cf. Pauling \cite{Pau60}) of MgS is 1.3 and that
of SiC is 0.7. Both compounds therefore show a significant ionic contribution to
their bonding. There
are obvious similarities of structure and bonding properties of SiC and MgS. On
the other hand, no such similarity exists between MgS and solid carbon. This 
suggests that SiC is much better suited as substrate for MgS precipitation
than carbon. We propose therefore that MgS grows on SiC grains.

A possible  core grain is any compound that has a lattice structure similar to the MgS crystalline structure and a higher condensation temperature. Among the
compounds possibly formed in the stellar outflows also CaS satisfies both
conditions. It has a crystal structure, that is the closest to that of MgS and
becomes stable at a more than 200 K higher temperature than MgS (Fig. 
\ref{FigStabLim}). The  abundance of Ca in a stellar wind, however, is 16 times
lower than the Mg abundance, which is not very favourable for its role as core
grain for MgS growth. Nevertheless, we will also check this possibility.

%% file: sect4.lt
\section{Models with core-mantle grains}
\label{WindModels}

We performed model calculations of dust formation in stellar outflows according
to the wind model described in Sect.~\ref{SectKinGr} to check the assumption of
the core-mantle growth scenario of MgS formation. Grain growth is calculated as described in Sect.~\ref{SectDustGrowthEq}.

\begin{figure}[t]

\includegraphics[width=\hsize]{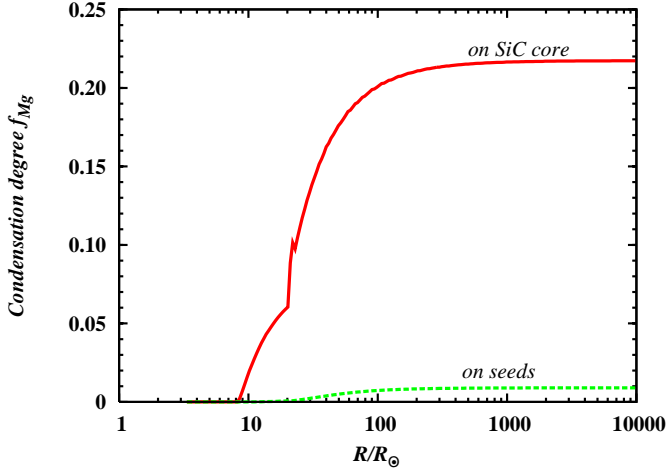}

\caption{Evolution of condensation degree of Mg with distance from
 the centre of the star for condensation of MgS (i) as mantle on top of a SiC
core and (ii) on seed particles (solid and dashed line, respectively).}
\label{FigCondDegree}
\end{figure}

First we check the model of MgS condensation on tiny seed nuclei. The existence of such seeds of 1~nm is taken for granted and the growth of MgS is calculated.
This results in grain sizes of only about $0.02\,\mu$m and a very low fraction
of Mg bound in dust, the condensation degree, of about $0.01$. This is in line with our analytical findings in Sect.~\ref{SectCondForGrowth}. The evolution of
grain radius and condensation degree for this kind of model are shown in 
Fig.~\ref{FigMgS_CaS} and Fig.~\ref{FigCondDegree}, respectively. The small amount of condensed MgS found in this case is obviously not enough to explain
the observed 30~$\mu$m emission in the spectra of carbon stars, underlining, again, the need for a core-mantle growth mechanism of MgS formation. 

Next we check an alternative growth scenario for MgS formation, in which first
CaS condenses on seed nuclei by the analogue of reaction (\ref{SiCfromH2S}), and then MgS grows as a mantle on this core. The model calculations show that the
size of CaS grains is much smaller than the required radius of $0.1~\mu$m. We
also tested the case, that MgS condensation is delayed by imposing the additional condition for the growth rate $J_{\rm gr,MgS} \ge 3 J_{\rm gr,CaS}$ to
gain additional time for CaS core growth in order to get bigger grains. This
artificial delay results, indeed, in bigger CaS cores, but also this does 
not help to increase the final size of the MgS mantle. The results of these calculations of the evolution of the MgS grain radius with distance from the
centre of a star for normal and delayed condensation on CaS are also shown in
Fig.~\ref{FigMgS_CaS}. The figure shows that the condensation of MgS on a CaS
core results in almost the same grain radius as the growth of MgS on seeds
grains. Both these formation mechanisms can not provide the amount of MgS dust
that is observed to condense in stellar outflows.

SiC grain growth, however, commences at higher temperature and proceeds faster
than CaS growth, so that at the onset of MgS formation the SiC core radius
reaches $\propto 0.07\,\mu$m as shown in Fig.~\ref{FigMgS_SiC}. The results of
MgS condensation on tiny seed nuclei are shown in Fig.~\ref{FigMgS_SiC} for
comparison. The SiC core is big enough to allow the formation of a considerable thickness of the MgS coating, of order 0.02~$\mu$m, which results in a final
grain size of $0.9~\mu$m. This value is very close to the required grain sizes
to explain the shape of the extinction feature in the spectra of carbon stars.
We assume in the calculation that SiC grains do not grow anymore as soon as the
formation of the MgS mantle begins. Figure~\ref{FigMgS_SiC} show that growth of
SiC grain with and without termination due to MgS mantle formation does not
change the grain size very much. Thus, a model with MgS mantle formation on SiC 
does not change noticeably the fraction of Si condensed in SiC dust, but results
in a much higher value of the degree of condensation of Mg in MgS as compared to
the case of growth on seeds. This is illustrated in Fig.~\ref{FigCondDegree}.

\begin{figure}[t]

\includegraphics[width=\hsize]{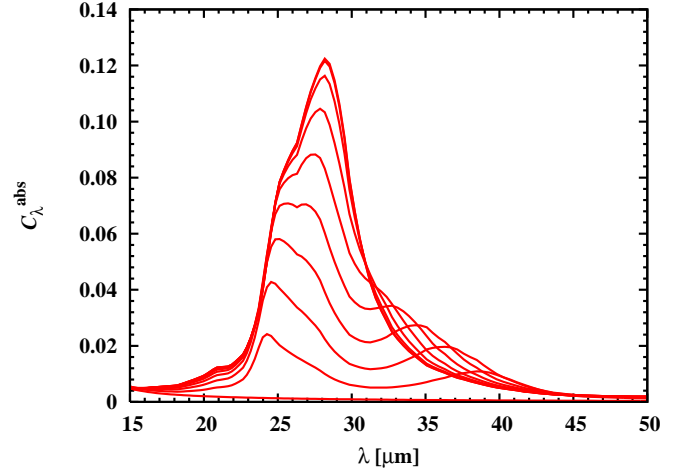}

\caption{Wavelength variation of the absorption efficiency $C_\lambda^{\rm abs}$
of spherical grains with SiC core of radius $r_{\rm c}$ coated with MgS mantles
of thickness $r_{\rm m}$ for a total grain radius $r_{\rm c}+r_{\rm m}$ of 
0.1~$\mu$m and for MgS-mantle thickness to total radius ratios from 0 to 1 in
steps of 0.1 (from bottom to top).}
\label{FigEqlComp}
\end{figure}

A similar test calculations for MgS coating formation on carbon grains also 
gave sufficient thicknesses of MgS layers. Basically, condensation of MgS on
carbon cores cannot be excluded, but from a physical point of view, MgS mantle
formation is more likely on a SiC core, since the bond lengths and bonding
properties of SiC are similar to that of MgS. Laboratory experiments would be
necessary to arrive at a final conclusion about the carriers of MgS mantles.

%% file: sect5.lt
\begin{figure}[t]

\includegraphics[width=\hsize]{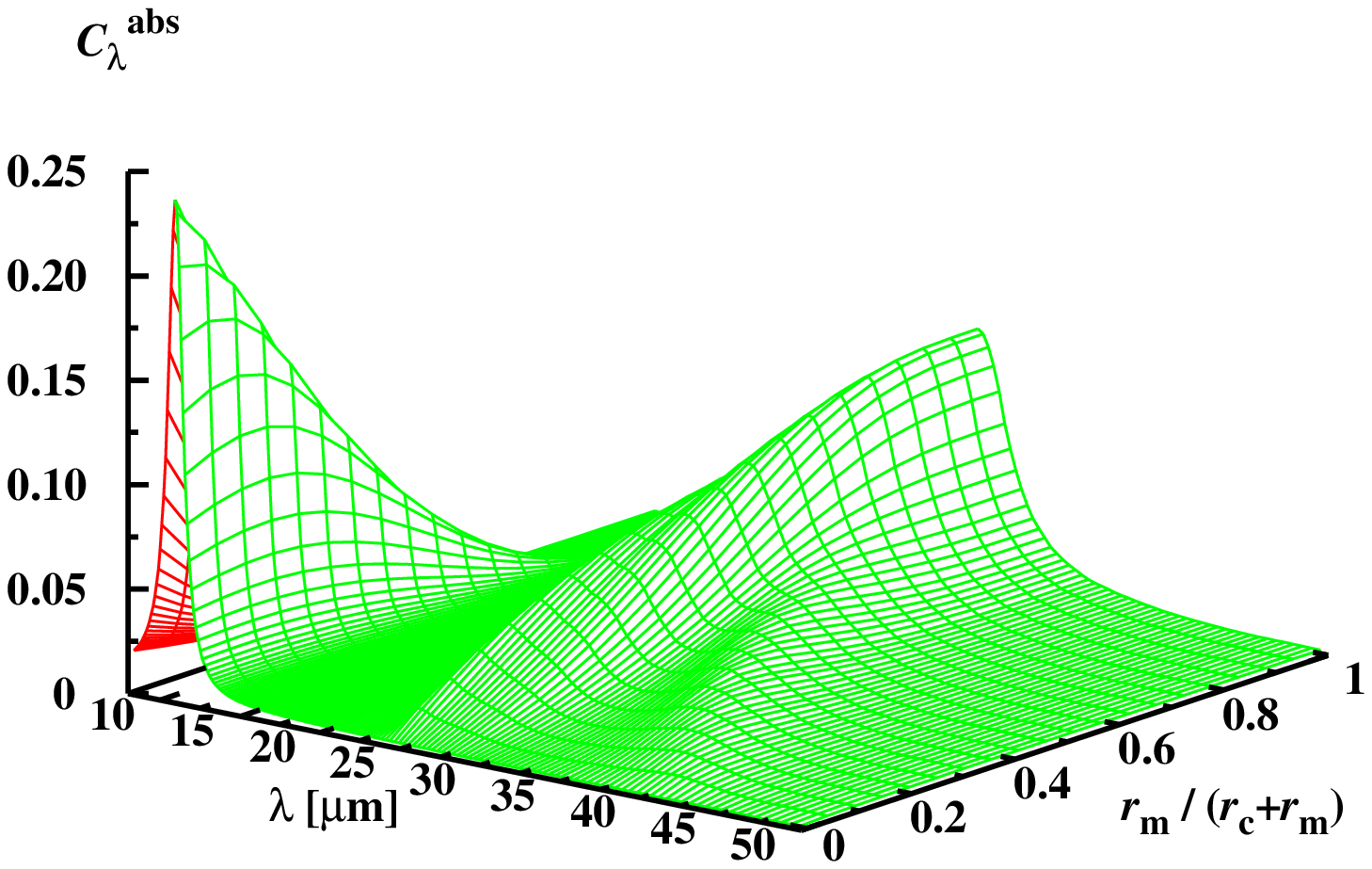}

\includegraphics[width=\hsize]{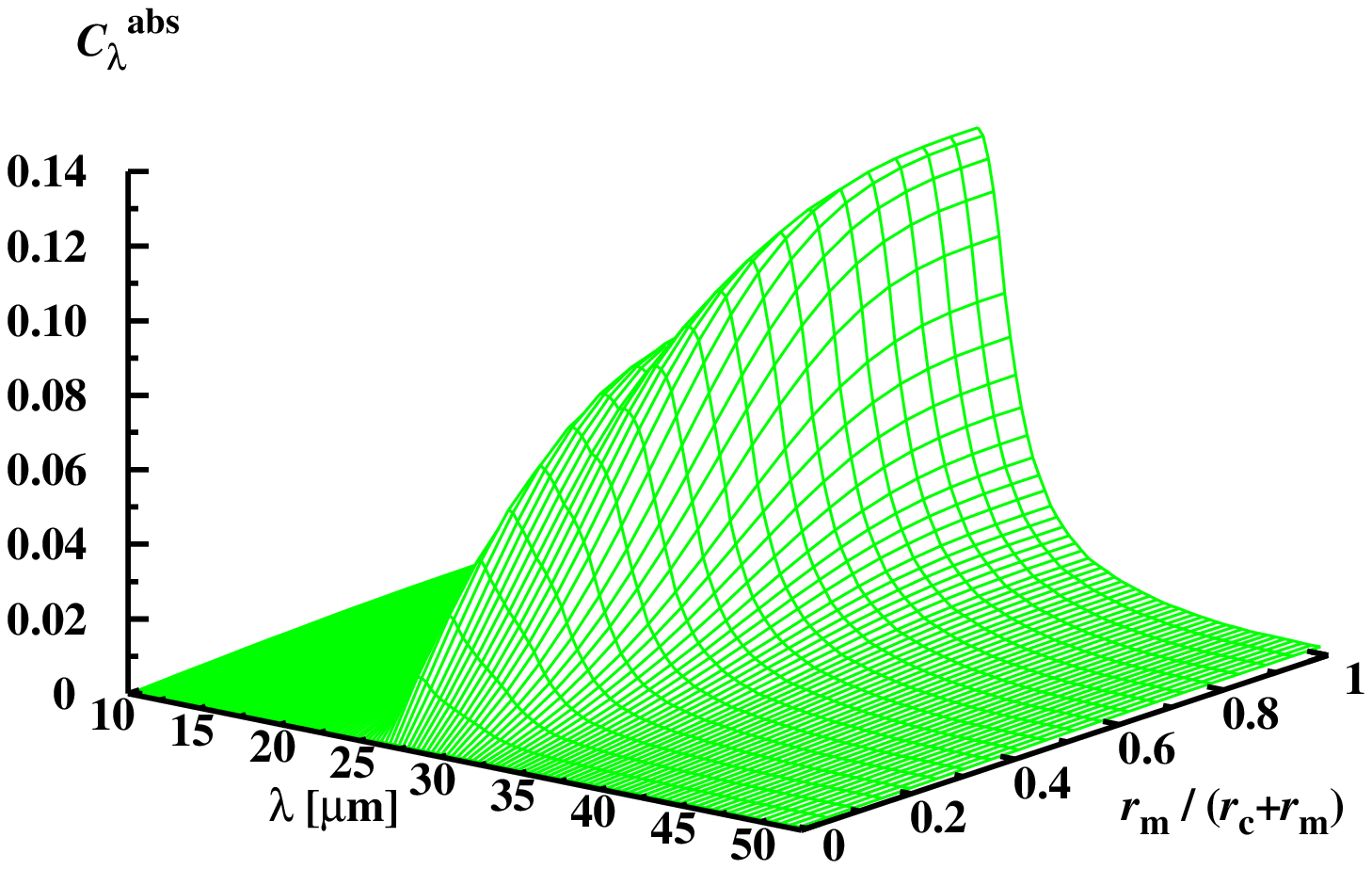}

\caption{Wavelength variation of the absorption efficiency $C_\lambda^{\rm abs}$
of spherical grains with SiC (top) or carbon (bottom) core and MgS mantle with total radius of 0.1~$\mu$m. The MgS mantles range in thickness from zero
to unity in steps of 0.05 for the ratio of mantle thickness $r_{\rm m}$ to total
radius $r_{\rm c}+r_{\rm m}$, where $r_{\rm m}$ is the core radius. The
extinction properties vary from that of pure SiC or carbon grains to that of pure
MgS grains. An extra extinction feature centred on 33 $\mu$m for MgS grains with
SiC core shows up at moderate mantle thickness. This is missing in case of carbon cores.}
\label{FigEqlComp3d}
\end{figure}

\section{Optical properties of grains with SiC core and MgS mantle}
\label{SectCoreManle}

In order to study the difference of optical properties of grains with coatings
of MgS either on a SiC core or on a carbon core we calculate the absorption
efficiency $C_\lambda^{\rm abs}$ (i.e. the ratio of absorption to geometrical
cross-section) in the small particle limit (cf. Bohren \& Huffman 
\cite{Boh83})
\begin{equation}
C_\lambda^{\rm abs} = 4x\,\textrm{Im} {\alpha\over 4\pi (r_{\rm c}+r_{\rm m})^3} 
\end{equation}
with
\begin{eqnarray}
\alpha&=&4\pi(r_{\rm c}+r_{\rm m})^3 \nonumber\\
&&\quad\displaystyle
{(r_{\rm c}+r_{\rm m})^3(\epsilon_2-1)(\epsilon_1+2\epsilon_2)
+ r_{\rm c}^3(2\epsilon_2+1)(\epsilon_1-\epsilon_2)\over
(r_{\rm c}+r_{\rm m})^3(\epsilon_2+2)(\epsilon_1+2\epsilon_2)
+ r_{\rm c}^3(2\epsilon_2-2)(\epsilon_1-\epsilon_2)  } \,.
\label{PolarCoat}
\end{eqnarray}
Here $r_{\rm c}$ is the core radius, $r_{\rm m}$ the thickness of the mantle and 
$\epsilon_1$ and $\epsilon_2$ are the complex dielectric functions of the core
and mantle material, respectively. $x$ is the size parameter $2\pi(r_{\rm c}+
r_{\rm m})/\lambda$. The small particle approximation $x\ll1$ is valid
in our case since we are interested in wavelengths from the region from 10 to 
40~$\mu$m that are much bigger than the size of circumstellar dust particles.

Optical constants for MgS are taken from Begemann et al. (\cite{Beg94}) for
Mg$_x$Fe$_{1-x}$S with $x=0.9$. For carbon we use data for the BE carbon dust
of the evaluation of the data of Colangeli et al. (\cite{Col93}) by Zubko et
al. (\cite{Zhu96}). For SiC we used the data for SiC from Laor \& Draine 
(\cite{Lao93}). Extinction efficiencies have been calculated for SiC grains
coated with MgS mantles and, for comparison, also for carbon grains with MgS
mantles.
 
Figure \ref{FigEqlComp} shows results for grains of total radius $r_{\rm c}+
r_{\rm m}$ of 0.1 $\mu$m and mantle thicknesses between 0 and $0.1~\mu$m, i.e.,
for grains ranging in composition from pure SiC to pure MgS grains. 
The results show the rather broad MgS absorption band centred around about 
26~$\mu$m, which in itself shows some structure. Additionally there appears a secondary peak centred around $\approx33\ \dots\ 38\ \mu$m for not too big mantle
thicknesses. Its position depends on the size ratio of core and mantle. This
feature is not present in the extinction efficiencies of MgS or SiC and results
from the particular run of the complex dielectric functions of the core and the
coating in this wavelength region\footnote{It has been checked if this results
from a Fr\"ohlich mode in coated particles at a frequency where the denominator
in Eq.~(\ref{PolarCoat}) vanishes, cf. the discussion of this effect in Bohren
\& Huffman (\cite{Boh83}), but this seems not to be the case}.

Figure \ref{FigEqlComp3d} shows the variation of the extinction coefficient 
with varying core to mantle size ratios -- both for MgS grains with
SiC and with carbon core -- in a different representation and an extended wavelength region that also covers the 11~$\mu$m feature of SiC. The total grain radius $r_{\rm c}+r_{\rm m}$, again, is 0.1 $\mu$m and the fraction
$r_{\rm m}/(r_{\rm c}+r_{\rm m})$ is varied between 0 and 1. In the upper part of the picture one recognises how the SiC feature disappears and the MgS feature appears as the composition of the grain varies from pure SiC to pure MgS. The extra feature peaking at about 33 \dots\ 38 $\mu$m, depending on the mantle
thickness, is clearly seen in case of coatings of moderate, but not too small
fraction of the total size. For thick coatings the extra feature disappears
together with the 11 $\mu$m feature from the SiC core. The feature is not
very distinct and will probably hardly be detectable in many cases. 

The particular feature peaking at about 33 \dots\ 38 $\mu$m is missing in the
extinction of carbon grains coated with MgS, as can be seen from the lower part
of  Fig.~\ref{FigEqlComp3d}, because of the completely different properties of the dielectric function. Therefore this extra feature can be considered as an
indicator of a silicon carbide core of MgS grains, if the presence of MgS is
indicated by its absorption band around 26 $\mu$m. Observationally it was found
by Volk et al. (\cite{Vol02}) that the so called `30\,$\mu$m feature' of carbon
stars shows in some cases some structure and seems to consist of two overlapping
features at 26 $\mu$m and 33$\mu$m. Our results for the extinction of 
SiC-core-MgS-mantle grains indicate, that one just sees that the MgS in
outflows from carbon stars grows as mantle on silicon carbide cores. If it is
not seen this probably does not mean that in this case MgS precipitates on carbon, but merely that the feature is not sufficiently well defined to be
unambiguously be detectable.

%% file: sect6.lt
\section{Concluding remarks}
\label{SectConcl}

In this paper the mechanism of MgS formation in stellar outflows of AGB stars is
studied. From some elementary considerations on the kinetics of MgS condensation
we estimate a critical mass-loss rate of about $5\times10^{-4}~\rm M_{\sun}\,
yr^{-1}$ for efficient MgS formation if this would condense as a separate dust
species via nucleation and subsequent growth. This value of the mass-loss rate is much higher than any observed mass-loss rate of AGB stars. Within the
observed range of mass-loss rates of AGB stars, however, the amount of MgS that
can condense on seed  particles in stellar winds is much too low to explain the
observed strong emission band from MgS in the infrared spectra of many carbon
stars.  

A model of heterogeneous growth of MgS on SiC precursor grains is proposed here 
as the most promising mechanism of MgS formation. Analysis of the equilibrium
chemistry of sulphur compounds in a carbon rich mixture shows that the formation
of MgS is  strongly coupled to the process of silicon carbide condensation.
If no SiC is formed in a stellar outflow, the sulphur is almost completely bound
in SiS molecules. Only after significant consumption of the abundant gas phase
SiS by SiC formation, $\rm H_2S$ is formed in abundance and reactions of this
with Mg makes possible MgS condensation. 

Our model calculations of MgS formation in stellar outflows show that only MgS
growth as mantle atop SiC cores, that condensed before the on-set of MgS
formation, results in sufficiently high degrees of condensation of Mg into MgS
of the order of $\sim0.2$ which are required that MgS may be clearly visible in
the infrared spectrum. Condensation of MgS on tiny seed particles would result
in a fraction of the Mg condensed in MgS dust of the order of only ~0.01.
This is much too small to be detectable.

Additionally, we performed calculations of the extinction properties of grains
with SiC core and MgS coating of various thicknesses. The presence of a SiC
core inside of MgS grains results within some range of core/mantle volume
ratios in a secondary peak near 33 \dots\ 35~$\mu$m in the broad emission band
associated with MgS, that seems to be observed in some spectra of AGB stars.
This feature is absent if MgS forms a mantle on carbon grains. We propose this
feature as an indicator for the presence of a silicon carbide core of MgS
grains, if the presence of MgS is indicated by its absorption band around 26
$\mu$m.

%% file: lit.lt
\begin{acknowledgements}
This work was supported by the Deutsche Forschungs\-gemeinschaft (DFG),
Sonderforschungsbereich 439 ``Galaxies in the Young Universe''. S. Z.
is supported in part also by the International Max-Planck Research School
(IMPRS) Heidelberg
\end{acknowledgements}


%% file: la.bbl
\begin{thebibliography}{}

\bibitem[2001]{All01}
Allende Prieto, C., Lambert, D.~L., \& Asplund, M. 2001, \apj, 556, L63

\bibitem[2002]{All02}
Allende Prieto, C., Lambert, D.~L., \& Asplund, M. 2002, \apj, 573, L137

\bibitem[2005]{Asp05}
Asplund, M., Grevesse, N., \& Sauval, A.~J. 2005, in Cosmic Abundances as
Records of Stellar Evolution and Nucleosynthesis, ASP Conference Series, 
Vol. 336, ed. T.~G. Barnes III, F.~N. Bash. (San
Francisco: Astronomical Society of the Pacific), p.25

\bibitem[1994]{Beg94}
Begemann, B., Dorschner, J., Henning, T., Mutschke, H., \& Thamm, E. 1994,
\apj, 423, L71

\bibitem[1983]{Boh83}
Bohren, C.~F., \& Huffman, D.~R. 1983, Absorption and Scattering
of Light by Small Particles. (New York: Wiley \& Sons) 

\bibitem[1988]{Bow88}
Bowen, G.~H. 1988, \apj, 329, 299

\bibitem[2007]{Cha07}
Chaplin, W.~J., Serenelli, A.~M., Basu, S., Elsworth, Y., New, R., \& Verner,
G.~A.\ 2007, \apj, 670, 872 

\bibitem[1998]{Chas98}
M. W. Chase: NIST-JANAF Thermochemical Tables. 4th edn. Journal of Physical and 
Chemical Reference Data, Monograph No. 9 (1998)

\bibitem[1993]{Col93}
Colangeli, L., Menella, V., Blanco, A. et al. 1993, \apj, 418, 435

\bibitem[1993]{Cox93}
Cox, P. 1993, in Astronomical Infrared Spectroscopy: Future Observational
Directions, ASP Conf. Ser., 41, 163

\bibitem[2003]{Dau03}
Daulton, T.~L., Bernatowicz, T.~J., Lewis, R.~S., Messenger, S., Stadermann,
F.~J., \& Amari, S. 2003, \gca, 67, 4743

\bibitem[2006]{Fer06}
Ferrarotti, A.~S., \& Gail, H.-P. 2006, \aap,   447, 553

\bibitem[1993]{Feu93}
Feuchtinger, M.~U., Dorfi, E.~A., \& H{\"o}fner, S. 1995, \aap, 273, 513

\bibitem[1991]{Fle91}
Fleischer, A.~J., Gauger, A., \& Sedlmayr, E. 1991, \aap, 242, L1

\bibitem[1992]{Fle92}
Fleischer, A.~J., Gauger, A., \& Sedlmayr, E. 1992, \aap, 266, 321

\bibitem[1981]{For81}
Forrest, W.~J., Houck, J.~R., \& McCarthy, J.~F. 1981, \apj, 248, 195

\bibitem[2003]{Gai03}
Gail, H.-P. 2003, In Astromineralogy, ed. Th. Hennig. (Heidelberg: Springer), p. 55

\bibitem[1986]{Gai86}
Gail H.-P., \& Sedlmayr E., 1986, \aap, 166, 225

\bibitem[1987]{Gai87}
Gail H.-P., \& Sedlmayr E., 1987, \aap, 171, 197

\bibitem[1985]{Goe85}
Goebel, J.~H., \& Moseley, S.~H., 1985, \apj, 290, L35

\bibitem[1998]{Gre98}
Grevesse, N., \& Sauval, A.~J. 1998, \ssr, 85, 161

\bibitem[2001]{Gri01}
Grishko, V.~I., Tereszchuk, K., Duley, W.~W., \& Bernath, P. 2001, \apj,
558, L129

\bibitem[1990]{Has90}
Hashimoto, A. 1990, Nature, 347, 53

\bibitem[1996]{Hen96}
Henning, T., \& Stognienko, R.\ 1996, \aap, 311, 291 

\bibitem[1997]{Hoe97}
H\"ofner, S., \& Dorfi, E.~A. 1997, \aap, 319, 648

\bibitem[1998]{Hoe98}
H\"ofner, S., J\o rgensen, U.~G., Loidl, R., \& Aringer, B. 1998, \aap, 340, 497

\bibitem[2002]{Hon02}
Hony, S., Waters, L.~B.~F.~M., \& Tielens, A.~G.~G.~M. 2002, \aap,390, 533

\bibitem[2004]{Hon04}
Hony, S., \& Bouwman, J 2004, \aap, 413, 981

\bibitem[2000]{Hri00}
Hrivnak, B.~J., Kwok, S., \& Kwok, S. 2000, \apj, 535, 275

\bibitem[2003]{Jeo03}
Jeong, K.~S., Winters, J.~M., Le~Bertre, T., \& Sedlmayr, E. 2003, \aap, 407,
191

\bibitem[1999]{Jia99}
Jiang, B.~W., Szczerba, R., \& Deguchi, S. 1999, \aap, 344, 918

\bibitem[2005]{Kim05} Kimura, Y., Kurumada, M., Tamura, K., Koike, C.,
Chihara, H., \& Kaito, C.\ 2005, \aap, 442, 507 

\bibitem[1993]{Lao93}
Laor, A., \& Draine, B.T. 1993, \apj, 402,441

\bibitem[1978]{Lat78}
Lattimer, J.~M., Schramm, D.~N., \& Grossman, L. 1978, \apj, 219, 230

\bibitem[2006]{Lag07}
Lagadec, E., Zijlstra, A.~A., Sloan, G.~C. and 11 coauthors 2007, \mnras, 376,
1270

\bibitem[2008]{Lei08}
Leisenring, J.~M., Markwick-Kemper, F., Sloan, G.~C. (2008) Effects on
metallicity on the chemical composition of carbon stars (preprint)

\bibitem[1985]{Nut85}
Nuth, J.~A., Moseley, H., Silverberg, R.~F., Goebel, J.~H., \& Moore, W.~J.
1985, \apj, 290, L41

\bibitem[1993]{Omo93}
Omont, A. 1993, in Astronomical Infrared Spectroscopy: Future Observational
Directions, ASP Conf. Ser., 41, 87

\bibitem[1995]{Omo95}
Omont, A., Moseley, S.~H, Cox, P. et al. 1995, \apj, 454, 819

\bibitem[2000]{Pap00}
Papoular, R. 2000, \aap, 362, L9

\bibitem[1960]{Pau60}
Pauling, L. 1960, The Nature of the Chemical Bond and the Structure of Molecules
and Crystals, 3rd. Ed. (Cornell University press)

\bibitem[1991]{Rou91}
Rouleau, F., \& Martin, P.~G. 1991, \apj, 377, 526

\bibitem[2006]{Slo06}
Sloan, G.~C., Kraemer, K.~E., Matsuura, M., Wood, P.~R., Price, S.~D., \&
Egan, M.~P. 2006, \apj, 645, 1118

\bibitem[1999]{Szc99}
Szczerba, R., Henning, Th., Volk, K., \& Kwok, S. 1999, \aap, 345, L39

\bibitem[1999]{vLo99}
van Loon, J.~T.,  Groenewegen, M.~A.~T., de Koter, A., Trams, N.~R., Waters,
L.~B.~F.~M.,  Zijlstra, A.~A., Whitelock, P.~A., \& Loup, C.\ 1999, 
\aap, 351, 559 

\bibitem[1993]{Vas93}
Vassiliadis, E., \& Wood, P.~R. 1993, \apj, 413, 641

\bibitem[2002]{Vol02}
Volk, K., Kwok, S., Herivnak, B.~J., \& Szczerba, R. 2002, \apj, 567, 412

\bibitem[1997]{Win97}
Winters, J.~M., Fleischer, A.~J., Le Bertre, T., \& Sedlmayr, E. 1997, \aap,
326, 305

\bibitem[2000]{Win00}
Winters, J.~M., Le Bertre, T., Jeong, K.~S., Helling, C., \& Sedlmayr, E. 2000,
\aap, 361, 641

\bibitem[1998]{Yam98}
Yamamura, I., de Jong, T., Justtanont, K., Cami, J., \& Waters, L.~B.~F.~M.
1998, \apss, 255, 351

\bibitem[1996]{Zhu96}
Zubko, V.~G., Menella, V., Colangeli, L., \& Busoletti, E. 1996, \mnras, 282, 1321

\bibitem[2006]{Zij06}
Zijlstra, A.~A., Matsuura, M., Wood, P.~R. and 13 coauthors 2006, \mnras, 365, 370,
1961

\end{thebibliography}
